\documentclass[a4paper,11pt]{article}
\pdfoutput=1 

\usepackage{jcappub}

\usepackage{lmodern}

\usepackage[T1]{fontenc}
\usepackage[utf8]{inputenc}
\usepackage{color}
\usepackage[usenames,dvipsnames]{xcolor}
\usepackage{amsmath}
\usepackage{amssymb}
\usepackage{graphicx}
\usepackage{esint}
\usepackage{placeins}

\usepackage{hyperref}
\definecolor{darkgreen}{rgb}{0.1,0.6,0.7}
\hypersetup{colorlinks = true,urlcolor=blue,citecolor=Blue}

\makeatletter

\let\jnl@style=\rm
\def\ref@jnl#1{{\jnl@style#1}}

\def\aj{\ref@jnl{AJ}}                   
\def\actaa{\ref@jnl{Acta Astron.}}      
\def\araa{\ref@jnl{ARA\&A}}             
\def\apj{\ref@jnl{ApJ}}                 
\def\apjl{\ref@jnl{ApJ}}                
\def\apjs{\ref@jnl{ApJS}}               
\def\ao{\ref@jnl{Appl.~Opt.}}           
\def\apss{\ref@jnl{Ap\&SS}}             
\def\aap{\ref@jnl{A\&A}}                
\def\aapr{\ref@jnl{A\&A~Rev.}}          
\def\aaps{\ref@jnl{A\&AS}}              
\def\azh{\ref@jnl{AZh}}                 
\def\baas{\ref@jnl{BAAS}}               
\def\bac{\ref@jnl{Bull. astr. Inst. Czechosl.}}
\def\caa{\ref@jnl{Chinese Astron. Astrophys.}}
\def\cjaa{\ref@jnl{Chinese J. Astron. Astrophys.}}
\def\icarus{\ref@jnl{Icarus}}           
\def\jcap{\ref@jnl{J. Cosmology Astropart. Phys.}}
\def\jrasc{\ref@jnl{JRASC}}             
\def\memras{\ref@jnl{MmRAS}}            
\def\mnras{\ref@jnl{MNRAS}}             
\def\na{\ref@jnl{New A}}                
\def\nar{\ref@jnl{New A Rev.}}          
\def\pra{\ref@jnl{Phys.~Rev.~A}}        
\def\prb{\ref@jnl{Phys.~Rev.~B}}        
\def\prc{\ref@jnl{Phys.~Rev.~C}}        
\def\prd{\ref@jnl{Phys.~Rev.~D}}        
\def\pre{\ref@jnl{Phys.~Rev.~E}}        
\def\prl{\ref@jnl{Phys.~Rev.~Lett.}}    
\def\pasa{\ref@jnl{PASA}}               
\def\pasp{\ref@jnl{PASP}}               
\def\pasj{\ref@jnl{PASJ}}               
\def\rmxaa{\ref@jnl{Rev. Mexicana Astron. Astrofis.}}%
\def\qjras{\ref@jnl{QJRAS}}             
\def\skytel{\ref@jnl{S\&T}}             
\def\solphys{\ref@jnl{Sol.~Phys.}}      
\def\sovast{\ref@jnl{Soviet~Ast.}}      
\def\ssr{\ref@jnl{Space~Sci.~Rev.}}     
\def\zap{\ref@jnl{ZAp}}                 
\def\nat{\ref@jnl{Nature}}              
\def\iaucirc{\ref@jnl{IAU~Circ.}}       
\def\aplett{\ref@jnl{Astrophys.~Lett.}} 
\def\apspr{\ref@jnl{Astrophys.~Space~Phys.~Res.}}
\def\bain{\ref@jnl{Bull.~Astron.~Inst.~Netherlands}} 
\def\fcp{\ref@jnl{Fund.~Cosmic~Phys.}}  
\def\gca{\ref@jnl{Geochim.~Cosmochim.~Acta}}   
\def\grl{\ref@jnl{Geophys.~Res.~Lett.}} 
\def\jcp{\ref@jnl{J.~Chem.~Phys.}}      
\def\jgr{\ref@jnl{J.~Geophys.~Res.}}    
\def\jqsrt{\ref@jnl{J.~Quant.~Spec.~Radiat.~Transf.}}
\def\memsai{\ref@jnl{Mem.~Soc.~Astron.~Italiana}}
\def\nphysa{\ref@jnl{Nucl.~Phys.~A}}   
\def\physrep{\ref@jnl{Phys.~Rep.}}   
\def\physscr{\ref@jnl{Phys.~Scr}}   
\def\planss{\ref@jnl{Planet.~Space~Sci.}}   
\def\procspie{\ref@jnl{Proc.~SPIE}}   

\newcommand{\mathd}{\ensuremath{\mathrm{d}}}

\newcommand{\both}{\ensuremath{\boldsymbol{\theta}}}

\newcommand{\fatt}{\ensuremath{\boldsymbol{x}}}

\newcommand{\fattheta}{\ensuremath{\boldsymbol{\theta}}}

\title{\boldmath Objective Bayesian analysis of neutrino masses and hierarchy}

\author[a]{Alan F. Heavens,}
\author[b]{Elena Sellentin}
\affiliation[a]{Imperial Centre for Inference and Cosmology (ICIC), Imperial College, Blackett Laboratory, Prince Consort Road, London SW7 2AZ, U.K.}
\affiliation[b]{Département de Physique Théorique, Université de Genève, 24 Quai Ernest-Ansermet, CH-1211 Genève, Switzerland}
\emailAdd{a.heavens@imperial.ac.uk, elena.sellentin@unige.ch}


\abstract{Given the precision of current neutrino data, priors still impact noticeably the constraints on neutrino masses and their hierarchy. To avoid our understanding of neutrinos being driven by prior assumptions, we construct a prior that is mathematically minimally informative.  
Using the constructed uninformative prior, we find that the normal hierarchy is favoured but with inconclusive posterior odds of 5.1:1. Better data is hence needed before the neutrino masses and their hierarchy can be well constrained. We find that the next decade of cosmological data should provide conclusive evidence if the normal hierarchy with negligible minimum mass is correct, and if the uncertainty in the sum of neutrino masses drops below 0.025 eV.  On the other hand, if neutrinos obey the inverted hierarchy, achieving strong evidence will be difficult with the same uncertainties. Our uninformative prior was constructed from principles of the Objective Bayesian approach. The prior is called a reference prior and is minimally informative in the specific sense that the information gain after collection of data is maximised. The prior is computed for the combination of neutrino oscillation data and cosmological data and still applies if the data improve.}

\begin{document}

\maketitle

\section{Introduction}
\label{firstpage} 
The precise masses $m_1, m_2$  and $m_3$ of neutrinos and the ordering of these masses are currently poorly determined features of particle physics.  Solar, atmospheric, accelerator and reactor neutrino experiments are able to determine mass-squared differences, and the so-called Mikheyev-Smirnov-Wolfenstein matter effect determines the ordering of two of the mass eigenstates, see e.g. \citet{MSW}. However, the absolute masses are currently relatively poorly constrained without recourse to cosmological data,  giving various 90\% credible upper limits for $m_1$  in the range $0.18-0.48$ eV, depending on the datasets used \citep{KZ,Chen}.  This leaves open both possibilities of $m_3$ being larger than $m_1$ and $m_2$ (the normal hierarchy, which we abbreviate by `NH'), and smaller than $m_1$ and $m_2$  (the inverted hierarchy;`IH').  

Current particle physics experiments are not able to make strong statements in favour of one ordering or the other \citep{Nufit,QianVogel,de Salas}. Cosmological observations add important information, since massive neutrinos alter, in a scale-dependent way, the growth of fluctuations. The principal sensitivity arises because the primordial fluctuations are erased on small scales by neutrino free-streaming, largely while the neutrinos are still relativistic.  This leads to a suppression of power at all but the lowest wavenumbers ($k<10^{-2}h$Mpc$^{-1}$) in the cosmic structures. 
In principle, such a suppression could be achieved by other sufficiently abundant relativistic free-streaming particles, but neutrinos are the particles preferred by the data \citep{SellentinDurrer}.
Since the proportion of the total fluctuation carried by the neutrinos increases as their total mass increases, the observed suppression of power allows us to infer model-dependent bounds on the sum of neutrino masses.  

Current data put an upper limit on the sum of the masses that is getting interestingly close to the minimum mass sums that are allowed by neutrino oscillations (0.06 and 0.1~eV for the NH and IH respectively).  For example, \cite{PD2015b} find an upper limit of 0.12 eV at 95\% confidence after having marginalised over other cosmological parameters from BOSS Ly$\alpha$ clustering and the cosmic microwave background. This result is similar to other studies using different probes and redshift ranges, such as those of  \cite{Cuesta,Giusarma}.

Previous determinations of the relative odds of the normal to the inverted hierarchy usually moderately favour the NH, depending on the dataset used. However, the data are not so constraining, with the effect that the prior chosen can still have a considerable influence on the results, and this is the focus of this paper.  

A frequentist view was taken by \cite{Capozzi}, but most analyses are Bayesian (e.g. \cite{Blennow,Gerbino,HS2016,SJPV,Vagnozzi,Long}), so priors are necessary.  In summary, \citet{HS2016} and \citet{Gerbino} found an odds ratio in favour of the normal hierarchy of 1.5 with a uniform prior for the lightest neutrino mass. \citet{Vagnozzi} found odds of 1.8 and 3.3 depending on whether conservative or aggressive mass sum limits were chosen. Such limits are again a form of imposing a prior. The shape \citet{Vagnozzi} chosen for their prior was uniform over the sum of neutrino masses.  In contrast to these approaches, \citet{SJPV} used the logarithm of the mass as a parameter to specify the prior, and advocated a more complex hierarchical model,  with normal priors in the logarithm of the masses, with the mean and variance of the normal being hyper-parameters.  As a consequence, \citet{SJPV} found much larger odds favouring the normal hierarchy by 42:1, and \citet{Gariazzo} similarly found that a logarithmic prior gave strong odds for the NH, and weak preference for the NH with linear priors. An even higher odds ratio was found by \citet{Long} where priors were set on the mass matrix itself. 

Clearly, the quoted odds ratios still depend on the priors chosen. This indicates that the data are not powerful enough to override the choice of prior. Answering the physical question of which hierarchy is preferred, still depends strongly on the information contributed by the chosen prior. We regard it however as important that physical conclusions depend on prior assumptions as little as possible. In this paper, we shall therefore explicitly construct a prior that \emph{minimizes} the information that the prior contributes to the analysis. We shall see that such an uninformative prior indeed leads to an uninformative odds ratio, meaning better data are needed to infer the neutrino mass hierarchy.

\section{The meaning of priors in Bayesian inferences}

In Bayesian analysis, there are broadly two schools of thought concerning appropriate priors.  In the subjective Bayesian approach, which has been taken to date in Bayesian neutrino mass analysis, one chooses the prior and states it, since any inferences will be dependent on the choice. Without previous data the aim is often to try to choose a `non-informative' prior, such that the posterior is driven by the likelihood and not the prior. One difficulty can be that what constitutes a non-informative prior may not be obvious, and apparently innocuous choices such as uniform priors can be extremely informative in some respects.  The alternative school of thought, called Objective Bayesianism, constructs a prior that is, in a mathematically well-defined sense, as uninformative as possible, on information-theoretic grounds. This includes arguments based on the Kullback-Leibler divergence, Haar measures, or the maximum entropy principle \citep{Bergerbook, Eaton, Fraser, Heath}.

The concept which applies to the situation at hand are so-called reference priors, introduced by \cite{Bernardo79}. A reference prior ensures that when the probability distribution of model parameters is updated from the prior to the posterior, the information gain is maximised.  The prior then has an obvious merit in that it is determined by a mathematical procedure, rather than a subjective view, and it maximises the influence of the data on the inference. One complication is that, in multi-parameter problems, the reference prior can depend on the ordering of parameters \citep{Bernardo2005}, so may not yield a unique maximally non-informative prior. However, a very nice feature of the combination of neutrino experiments used in this paper is that the ordering of the parameters is irrelevant, and there is, in this case, a unique reference prior.  This prior is one of the main results of this paper. The approach is outlined in the next section; for a full discussion of reference priors the reader is directed to \cite{Bernardo2005,Berger}.  The other main results are the ensuing posteriors on the neutrino masses, and the resultant posterior odds of the two possible neutrino hierarchies (where with current data none is strongly preferred).

\section{Least-informative priors for a given experiment}

To make precise the notion of a `least-informative prior', consider a model $M$ with parameters $\fattheta$, which can generate data ${\fatt}$. The theory of uninformative priors has progressed in the past thirty years, and the concept which applies to the problem of infering the neutrino mass hierachy is the theory of `reference priors'. What is nowadays understood as a reference prior is a prior that maximises the expected information gain caused by the arrival of the data $\fatt$. In other words, a reference prior maximizes the missing information, in the absence of data. The concept of missing information is thereby central: it defines what is mathematically meant when we request an `uninformative' prior.

For a reference prior, the missing information is assessed as follows. Let there be an experiment, which aims to constrain parameters $\fattheta$. Being Bayesian, the experiment is only conducted once, and procures a data set $\fatt \sim p(\fatt|\fattheta)$. The likelihood then quantifies how tightly a single outcome of the experiment constrains the parameters $\fattheta$. In contrast, the missing information before knowing the parameter precisely depends additionally on the prior: if the prior is sharply peaked, it contributes a lot of information. To quantify the missing information, reference prior theory uses $k$ hypothetical repetitions of the experiment, as these would finally overwrite any sensible prior choice. For $k \to \infty$ repetitions, the parameter is then as precisely constrained as possible by the \emph{data}, conditional on the assumed experimental setup.  Note that the experiment does not need to be repeated in reality -- the hypothetical repetition is only necessary to quantify the missing information.  The reference prior is then the prior which maximizes this missing information.

Denoting the data generated by $k$ repetitions of the experiment as $\fatt^k$, the mutual information between a prior $\pi(\both)$ and posterior $p(\fattheta|\fatt^k)$  is \citep{Berger}
\begin{equation}
I(\fattheta,\fatt^k) = \int p(\fatt^k) \left[ \int p(\fattheta | \fatt^k) \ln \frac{p(\fattheta | \fatt^k)}{\pi(\fattheta)} d\fattheta\right]\,d\fatt^k,
\label{info}
\end{equation}
where we recognise the inner integral as the Kullback-Leibler divergence between the posterior $p(\fattheta | \fatt^k)$ and the prior $\pi(\fattheta)$. The outer integral runs over realizations of the data, meaning the information here defined is an \emph{expected} information. Maximising the missing information yields the on average least-informative prior $\pi(\fattheta)$.

The solution for the least-informative prior $\pi(\fattheta)$ amongst a class of candidate priors given the information measure eq.~\ref{info}, is \cite{Bernardo2005,Berger}
\begin{eqnarray}
& \pi(\fattheta)  = \lim_{k \to \infty }\frac{f_k(\fattheta)}{f_k(\fattheta_0)},\\
& f_k(\fattheta)  = \exp\left[\int p(\fatt^k | \fattheta) \ln p^*(\fattheta | \fatt^k)\,\mathd\fatt^k \right].
\label{reffi}
\end{eqnarray}
where $\fattheta_0$ is a point in the parameter space, and $p^*(\fattheta | \fatt^k)$ is the posterior with a fiducial choice of a prior $\pi^*(\fattheta)$, and the limit is loose in that it permits improper priors. The choice of the fiducial point $\fattheta_0$ and the fiducial prior $\pi^*$ can facilitate the computation, but (given the existence of the resulting reference prior and certain regularity conditions) do not influence the outcome. Typically, algorithmic sampling procedures are employed to solve eq.~\ref{reffi}. See \cite{Berger} for regularity conditions and proofs.

However, of particular convenience is that for likelihoods which already achieved asymptotic normality (roughly: a precision experiment), the reference priors for single parameters converge to a prior class which can be more easily computed, namely the so-called Jeffreys priors. Certain likelihoods may of course never achieve asymptotic normality. This is for example possible in multi-parameter problems in cases where the data cannot lift parameter degeneracies, see e.g.~\citep{DALI,DALII}. The reference prior which is least-informative according to \ref{info} does then not coincide with the Jeffreys prior.

The Jeffreys prior in one dimension is defined as
\begin{equation}
\pi_J(\theta) \propto \sqrt{ -\mathbb{E}\left[ \frac{d^2\ln p(\fatt|\theta)}{d\theta^2}\right]  },
\label{jeffi}
\end{equation}
where the expectation operator $\mathbb{E}$ is again the integral over the data, see e.g.~\cite{Bernardo2005, Jeffreys}. Note that the statistical definition of the Jeffreys prior is more general and also a deeper concept than what is usually called a `Jeffreys prior' in astronomy (where the astronomical `Jeffreys prior' is used to denote the prior $\pi \propto 1/\theta$). The general Jeffreys prior hence uses as information measure the Fisher information 
\begin{equation}
F = -\mathbb{E}\left[\frac{d^2\ln p(\fatt|\theta)}{d\theta^2}\right],
\label{Fish}
\end{equation}
rather than the mutual information of eq.~\ref{info}. 

For multi-parameter models with a likelihood function $p(\fatt|\fattheta)$, the reference prior is found sequentially, by first fixing all but one parameter and finding the reference prior $\pi(\theta_1 | \fattheta_{i\ne 1})$ for the remaining parameter $\theta_1$ by treating it as a single parameter problem.  Assuming the resulting prior is proper, then $\theta_1$ may be integrated out, and the likelihood is 
\begin{equation}
p(\fatt | \fattheta_{i\ne 1}) = \int p(\fatt | \fattheta)\pi(\theta_1)d\theta_1.
\label{marg}
\end{equation}
The priors are then obtained sequentially by repeating this procedure, and the final prior is $\pi(\fattheta) = \pi(\theta_n)\pi(\theta_{n-1}|\theta_n)\ldots \pi(\theta_1 | \fattheta_{i\ne 1})$.  Note that in some cases the prior obtained this way depends on the ordering of the parameters, and it is then conventional to order by the perceived importance of the parameters.  However, if the likelihood is separable into subsets of the data $\fatt$, i.e. $\prod_i p(\fatt_i | \fattheta_i)$, as in this case, then $\prod_{i\ne 1} p(\fatt_i | \fattheta_i)$ may be taken outside the integral in eq.(\ref{marg}), and repeated application of this principle shows that the reference prior is the product of the individual 1D reference priors, and is unique.  

A reference prior constructed either via \ref{reffi}, or if appropriate via \ref{jeffi}, has multiple attractive features. Firstly, it minimizes on average the information that the prior contributes to the measurement. Secondly, reference priors satisfy multiple sanity criteria. For example, reference priors are preserved when compressing the data onto sufficient summary statistics. Furthermore, reference priors are preserved when conducting independent repetitions of a measurement. This means reference priors do not tip from uninformative to informative with increasing sample size, as many other priors do which are a function of the number of data points \citep{SunBerger}. Reference priors are also consistent under invertible reparametrizations and they furthermore hold even if the initially accessible parameter space is retrospectively constrained to a compact subset \cite{Bernardo2005,Berger}. The latter property ensures that reference priors consistently handle unphysical regions\footnote{This can intuitively be understood by noting that the distribution function in eq.~\ref{info} will not generate data in unphysical regions, although the actual proof is somewhat more involved \citep{Berger}.}. 

Reference priors and Jeffreys priors are sometimes criticized by pure statisticians for depending on the experimental setup. This dependence ensues since the priors are computed from either infinite repetitions of the data, or from expectation values gained from integrating over the distribution of the data. For a physicist, this dependence on the experimental setup is however natural: different experiments will procure different data sets, and different data sets will overwrite the same prior with different speeds. It is therefore natural that a least-informative prior depends on the measurement design.

In the following sections we construct the reference prior for the investigated set of neutrino experiments, and use it to determine the neutrino mass posteriors, and the odds of the two neutrino hierarchies.

\subsection{Parametrisation of neutrino experiments}

The mutual information is parameter-independent, which means that the reference prior is also independent of parametrisation. This is convenient, because in order to constrain the neutrino mass hierarchy, it is most natural to parametrise the model directly with the neutrino masses $m_1, m_2, m_3$. In contrast, to compute the reference prior, it is simpler to start from the quantities that the experiments are actually sensitive to, namely the mass splittings and the sum of neutrino masses. We therefore order the masses according to  $m_{\rm H}\ge m_{\rm M}\ge m_{\rm L}$, where the subscripts H, M, L abbreviate `high, medium, low'. The hierarchy is then 
normal if $m^2_{\rm H}-m^2_{\rm M} > m^2_{\rm M}-m^2_{\rm L}$, and inverted otherwise.

The quantities that are constrained or measured by experiments are then
\begin{eqnarray}
\phi &\equiv& m_{\rm M}^2-m_{\rm L}^2 \quad {\rm Normal}\nonumber\\
&\equiv& \ m_{\rm H}^2-m_{\rm M}^2\quad {\rm Inverted}\nonumber\\
\psi &\equiv &m_{\rm H}^2-\frac{1}{2}(m_{\rm M}^2+m_{\rm L}^2)\quad  {\rm Normal} \nonumber\\
&\equiv &\frac{1}{2}(m_{\rm H}^2+m_{\rm M}^2)-m_{\rm L}^2 \quad {\rm Inverted}\nonumber\\
\Sigma &\equiv& m_{\rm H}+m_{\rm M}+m_{\rm L}. 
\label{pars}
\end{eqnarray}
The parameters $\phi$ and $\psi$ are the (squared) mass splittings, and the parameter $\Sigma$ is the sum of neutrino masses. For further discussion, see \cite{Long}. These parameters are constrained by solar experiments, a combination of long baseline (LBL) accelerator and reactor experiments, and cosmological neutrino experiments respectively.  Using the parametrisation in \cite{Long}, current constraints are \citep{Nufit,Capozzi,Long, Nulims, PD2015b,Cuesta}
\begin{eqnarray}
m_{\rm M}^2-m_{\rm L}^2 &=& (7.50 \pm 0.18) \times 10^{-5} {\rm\ eV}^2\ {\rm (Normal)}\nonumber\\
m_{\rm H}^2-m_{\rm M}^2 &=& (7.50 \pm 0.18) \times 10^{-5} {\rm\ eV}^2 \ {\rm (Inverted)}\nonumber\\
m_{\rm H}^2-(m_{\rm M}^2+m_{\rm L}^2)/2 &=& (2.524 \pm 0.04) \times 10^{-3} {\rm\ eV}^2 \ {\rm (Normal)}\nonumber\\
(m_{\rm H}^2+m_{\rm M}^2)/2-m_{\rm L}^2 &=& (2.514 \pm 0.04) \times 10^{-3} {\rm\ eV}^2 \ {\rm (Inverted)}\nonumber\\
m_{\rm L}+m_{\rm M}+m_{\rm H} &<& 0.12 {\rm\ eV} \qquad {\rm 95\%\ credible\ region}.
\label{data}
\end{eqnarray}
To construct the least informative prior, we first notice that the constraints above are means and their respective standard deviations, which is justified because the posteriors for the terrestrial experiments are already very close to Gaussian, see also \citep{Bergstrom,Nufit}. This means asymptotic posterior normality has already set in. Additionally, $\phi$ and $\psi$ are well determined, meaning the constraints in eq.~\ref{data} are insensitive with respect to priors potentially assumed in the original publications (if they were Bayesian analyses).  The cosmological probability of the neutrino mass sum is also approximately a Gaussian truncated at zero, but evidently not well measured yet \citep{Cuesta}.  To compute the reference priors for $\phi,\psi$ and $\Sigma$, we additionally note that the quoted experiments are independent of each other: especially the terrestrial and solar constraints on $\phi$ and $\psi$ were computed from data which are independent of cosmological data which constrain the sum of neutrino masses. The joint data set of terrestrial, solar and cosmological data therefore factorizes, and hence so does the least informative reference prior. We can therefore compute a total reference prior for the neutrino hierarchy, which is the product of the reference priors for $\phi,\psi,\Sigma$. We will then transform the reference prior for $\phi,\psi,\Sigma$ onto the reference prior for $m_{\rm L}, m_{\rm M}, m_{\rm H}$.

To compute the reference priors for $\phi,\psi,\Sigma$ we exploit that asymptotic posterior normality has already set in for these paramerers. This means we do not need to solve eq.~\ref{reffi} via a sampling algorithm, but can instead take the short-cut and compute the Jeffreys prior $\pi_J(\theta) \propto \sqrt{F(\theta)}$ from eq.~\ref{jeffi}. The Fisher information is given by eq.~\ref{Fish}. The Gaussian likelihoods are given by
\begin{equation}
L(\hat{\mu}|\theta) = \frac{1}{\sqrt{2\pi \sigma^2(\theta)}} \exp\left( - \frac{1}{2} \frac{\left[\hat{\mu}-\mu(\theta)\right]^2}{\sigma^2(\theta)} \right).
\end{equation}
where $\theta$ is any of the three parameters $\phi,\psi,\Sigma$ and $\hat{\mu}$ is the summary statistic of the data from which the parameters are estimated (here the estimated mean). The Fisher information is then
\begin{equation}
F(\theta) = \frac{\left[\partial_\theta \mu(\theta)\right]^2}{\sigma^2(\theta)} + 2 \frac{\left[\partial_\theta \sigma(\theta)\right]^2}{\sigma^2(\theta)}.
\label{LongFish}
\end{equation}
How this Fisher information now scales with the parameters determines the shape of the priors. We see that for $\theta$ being any of $\phi,\psi,\Sigma$, the likelihoods are already Gaussian, hence the parameters are effectively linear over the domain of relevance and we hence have $\mu(\theta) \propto \theta$ in all three cases. Moreover, we know from the measurement designs that the standard deviations $\sigma(\theta)$ are independent of $\theta$ over a wide range. For the cosmological measurements of $\Sigma$, this is because the covariance matrices are there taken to be cosmology independent. For the mass splittings $\phi$ and $\psi$, the independence of the noise is somewhat harder to see, but arises because of the following reasoning.

For a given neutrino oscillation experiment, the sensitivity $S$ varies periodically with the difference in the squared masses of two neutrino mass eigenstates $i,j$, according to
\begin{equation}
S \propto \sin^2\left[ \alpha \frac{(m_i^2 - m_j^2) L}{E_\nu}\right]
\end{equation}
where $L$ is the baseline, $E_\nu$ the neutrino energy, and $\alpha$ is a constant, containing the speed of light, etc. The sensitivity does however not depend on the absolute masses. We are interested in the prior on $\phi$ and $\psi$ over the range where the likelihood is high.  Ideally $L$ is optimised such that the angle $\alpha\psi L/(4 E_\nu) \sim \pi/2$,
\cite{Timmons} with the result that the expected error is sensitive to the mass splittings $\psi$ or $\phi$ only when the shift in mass difference changes the angle by $\mathcal{O}(1)$. This is more than an order of magnitude larger than the width of the likelihood. Therefore, $\phi$ and $\psi$ are indeed effectively linear parameters over the width of the likelihood, and the measurement uncertainty is constant over the still admissible range for $\phi$ and $\psi$.  

We therefore see that given the current measurement precisions, all three parameters $\phi,\psi,\Sigma$ can effectively be treated as linear and the noise can be treated as constant. The Fisher information \ref{LongFish} is then constant, since the derivative $\partial_\theta \sigma(\theta)$ vanishes, and since the derivative $\partial_\theta \mu(\theta)$ is effectively constant over the domain of high likelihood. Because of the Fisher information being constant, we arrive at uniform Jeffreys priors for all three parameters
\begin{equation}
\pi_J(\phi) = {\rm const.},\hspace{12pt}
\pi_J(\psi) = {\rm const.},\hspace{12pt}
\pi_J(\Sigma) = {\rm const.}
\end{equation}
Let us shortly discuss the approximations in this derivation. These priors were derived using that the parameters are effectively linear, due to the constraining power of the data. In order for this to be not only approximately but rather strictly true, there should be translational invariance of the likelihood also for \emph{less} constraining data. This translational invariance is meant in the sense that the likelihood only shifts as a function of each $\theta$ but does not change shape or width. For current data, assuming linear parameters is an excellent approximation, and has been studied especially for the cosmological experiment, where Fisher matrix analyses \citep{Takeuchi,Allison} have shown constant Fisher information as a function of the neutrino masses. For future data, the precision of this approximation will improve further. Note that although cosmological neutrino results are presented truncated at zero mass sum, these should be interpreted as a likelihood multiplied by a hard theoretical prior truncated at zero. The likelihood itself can extend to negative mass sum.  For example, an estimator of the mass sum is not required to be positive.  We make the assumption that the likelihood itself is translation-invariant, consistent with the Fisher analyses of \citep{Takeuchi,Allison}, in which case the Jeffreys prior is constant, regardless of its shape.

In addition to noting that truncating the distribution for the neutrino masses at $\Sigma=0$ does not affect the least informative reference prior, there are some regions of ($\phi,\psi,\Sigma$) space that are unphysical, since they correspond to negative values of one or more of the masses or mass-squared terms. In conjunction with later imposing an upper limit on the sum of masses, this can again be seen as a restriction of the parameter space. The reference prior handles this, since it can always be retrospectively constrained to a compact subspace. This can be proven, see \cite{Berger}, but in principle arises because independently gained information is additive, and restricting the prior to physically meaningful parameter ranges simply adds physical information to the missing information as assessed by the reference prior.

Having thus determined the reference prior in the ($\phi,\psi,\Sigma$) parametrisation, in the next section we use the invariance of the mutual information to parametrisation (and hence consistency of the reference prior) to find the prior in the ($m_{\rm L},m_{\rm M},m_{\rm H}$) parametrisation. From there, we will compute the posterior mass distributions and the marginal evidence for the two hierarchies.

\section{Construction and interpretation of the neutrino mass prior}
\label{ThePrior}
To compute the posterior odds of the two possible neutrino hierarchies, we take the uniform reference priors in $\phi,\psi,\Sigma$, and transform to $m_{\rm L},m_{\rm M},m_{\rm H}$. This produces the Jacobian 
\begin{eqnarray}
J(m_{\rm L},m_{\rm M},m_{\rm H}) & = & \left|\left|\frac{\partial(\phi,\psi,\Sigma)}{\partial (m_{\rm L},m_{\rm M},m_{\rm H})} \right|\right| \nonumber\\
&= & 4(m_{\rm L} m_{\rm M} + m_{\rm L} m_{\rm H} + m_{\rm M} m_{\rm H}),
\label{refprior}
\end{eqnarray}
for both hierarchies. $J$ is
therefore proportional to the resulting non-informative reference prior in the space of neutrino masses $m_i$. This prior makes immediate sense. Recall that the prior \ref{refprior} had been constructed from the information which an experiment misses on the parameters.

\begin{figure}
\includegraphics[width=\textwidth]{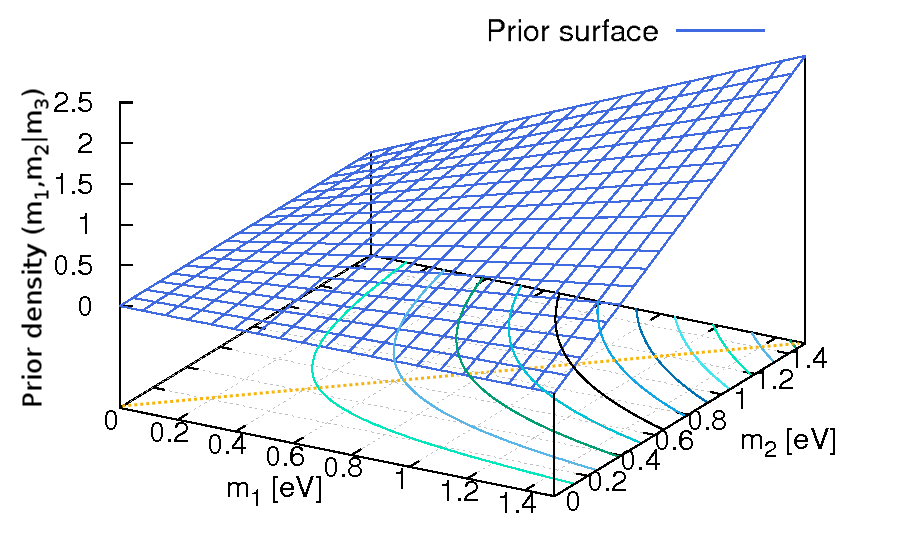}
\caption{Surface plot of the on average least informative prior density for the neutrino masses (keeping $m_3 = 0.01 {\rm eV}$ fixed as an example for illustration. This simply determines the intersection with the $z$-axis; the $z$-axis can be rescaled for other values of $m_3$). This prior was constructed from the reference priors for the mass splittings and the sum of neutrino masses, see eq.~\ref{refprior}. The prior correctly reflects that a priori, a low neutrino mass sum is more difficult to detect (low prior density to the bottom left). It also reflects that keeping one of the three neutrino masses fixed, it is most difficult to detect a neutrino mass splitting if the two other neutrinos have identical masses (minimal prior value along the line $m_1 = m_2$, indicated in yellow dashes). The coloured lines indicate isocontours.}
\label{PlotMassPrior}
\end{figure}

We now see that the prior is zero, if all neutrino masses are zero. The prior thereby correctly reflects that if the neutrino masses were all zero, then no experiment searching for the masses could ever detect them.

Keeping one of the three masses fixed ($m_{\rm L}$ say), the prior furthermore takes a minimal value if the two other masses coincide ($m_{\rm M} = m_{\rm H}$ in our example). This reflects that an experiment which searches for a mass-splitting between $m_{\rm M}$ and $m_{\rm H}$ will then fail, since there is then no mass splitting to be detected. 

The prior also increases with increasing masses, which reflects that it is a priori easier to detect the sum of the neutrino masses if it is large. This again makes immediate sense, as highly massive neutrinos will wash out cosmological structures efficiently. The prior hence directly reflects why cosmology is able to yield a sequence of increasingly constraining \emph{upper} bounds. Moreover, the prior increases with the sum of neutrino masses only in a polynomial fashion. This directly implies that it is a sound prior which can be overwritten by the data: for increasingly constraining data, the likelihood will turn towards Gaussian, thereby acquiring exponentially decreasing tails. As an exponential times a polynomial converges, this directly illustrates that the posterior will converge.

Comparing the prior \ref{refprior} to the often chosen priors that are flat in mass or proportional to $1/m$, we see a conceptual difference:  these priors describe a prior belief that a value for the masses or their logarithm is not preferred. The reference prior eq.~\ref{refprior} instead reflects whether the \emph{experimental setup} is designed to be more sensitive with respect to certain combinations of the values for the masses or not. It constructs the prior which is on average most easily overwritten by data. If we would reinterpret the $1/m$ prior as a reference prior, we see that it puts maximal weight on zero masses. This would imply an experiment is a priori most likely to detect the neutrino masses if they were zero (which is of course a contradiction in itself).

\section{Results: neutrino masses and hierarchy}

\subsection{Mass posteriors}
\begin{figure}
\includegraphics[width=\textwidth]{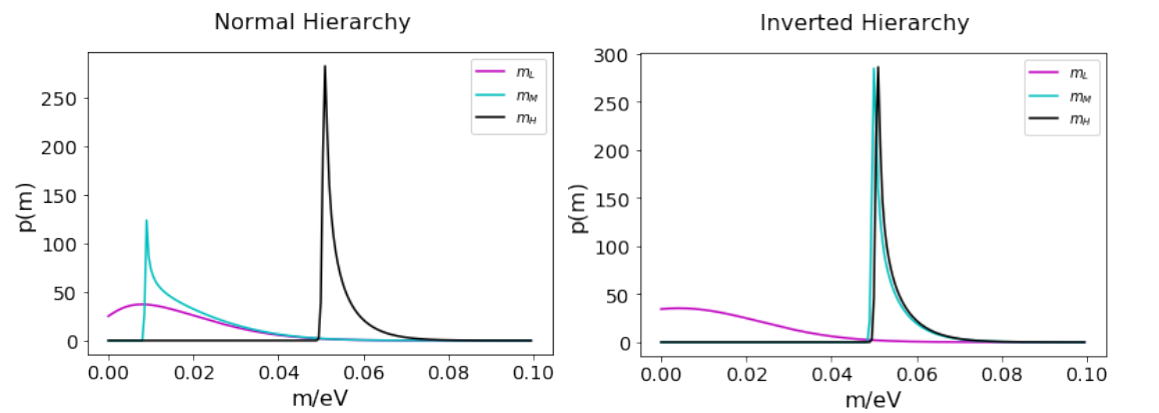}
\caption{Posteriors for masses for the normal Hierarchy (left), and the Inverted Hierarchy (right).}
\label{MassDistNH}
\end{figure}

Using the constructed least informative prior, eq.~\ref{refprior}, we show in figure \ref{MassDistNH} the posteriors for the neutrino masses for the two hierarchies.  As expected, for the inverted hierarchy, the two most massive states are tightly constrained by the accelerator/reactor result and the cosmological upper bound.  If we wish, we can marginalise over the model (NH or IH), in which case the posteriors become an average of these posteriors, weighted with the Bayes factor.  The most massive neutrino state is then tightly constrained, regardless of hierarchy.

\subsection{Marginal Likelihood for the hierarchies}

Having determined the reference prior, constructed to be non-informative for the neutrino masses, we now use it to compute the posterior probability of the normal hierarchy and inverted hierarchy. This model comparison is achieved with the Bayesian evidence, or marginal likelihood.  The relative probability of the two models (posterior odds) is 
\begin{equation}
\frac{p({\rm NH | \fatt})}{p({\rm IH | \fatt})} = \frac{\pi({\rm NH})}{\pi({\rm IH})} \frac{\int p(\fatt |\fattheta, {\rm NH}) \pi(\fattheta|{\rm NH}) \, \mathd\fattheta}{\int p(\fatt |\fattheta, {\rm IH}) \pi(\fattheta|{\rm IH}) \, \mathd\fattheta}.
\end{equation}
The first term is the prior model ratio, which we shall take to be unity to express no prior model preference.  The second term is the Bayes factor, being the ratio of the evidences, $B = Z_{\rm NH}/Z_{\rm IH}$.

The likelihood is given by the product of the three Gaussians for $\phi,\psi$ and $\Sigma$, or their transformed distributions for $m_{\rm L}, m_{\rm M}, m_{\rm H}$. They are combined with the prior \ref{refprior} such that we can compute the marginal likelihoods for NH and IH respectively:
\begin{equation}
Z_{\rm NH,IH} =A_{\rm NH,IH}\,\int \,dm_{\rm L} dm_{\rm M} dm_{\rm H}  \, J(m_{\rm L},m_{\rm M},m_{\rm H})   
 {\cal N}_{\phi}(\mu_{\rm s}, \sigma^2_{\rm s}) \, {\cal N}_{\psi}(\mu_{\rm a}, \sigma_{\rm a}^2)
 \, {\cal N}_{\Sigma}(\mu_{\Sigma}, \sigma^2_{\Sigma}),  
\end{equation}
with obvious integration limits. $A_{\rm NH,IH}$ are constants and we insert the appropriate functions of $m_i$ from Eq.~(\ref{pars}).  ${\cal N}_x(\mu,\sigma^2) = (2\pi \sigma^2)^{-1/2}\exp[-(x-\mu)^2/(2\sigma^2)]$ is the normal distribution, and the appropriate means and variances for the solar (s) and LBL accelerator/reactor (a) experiments are obtained from Eq.~(\ref{data}). For the cosmological data, we take $\mu_\Sigma=0$, and $\sigma_\Sigma \simeq 0.06$ eV \citep{PD2015b}.  The cosmological term is truncated at $\Sigma=0$. Note that these likelihoods each have a normalisation, which is common for all except the LBL/reactor results, for which the IH peak likelihood is lower by $\Delta\chi^2=0.83$ \citep{Nufit}, so the NH/IH Bayes factor is increased by $A_{\rm NH}/A_{\rm IH} = \exp(0.415)\sim 1.5$.

\begin{figure}
\centering
\includegraphics[width=8cm]{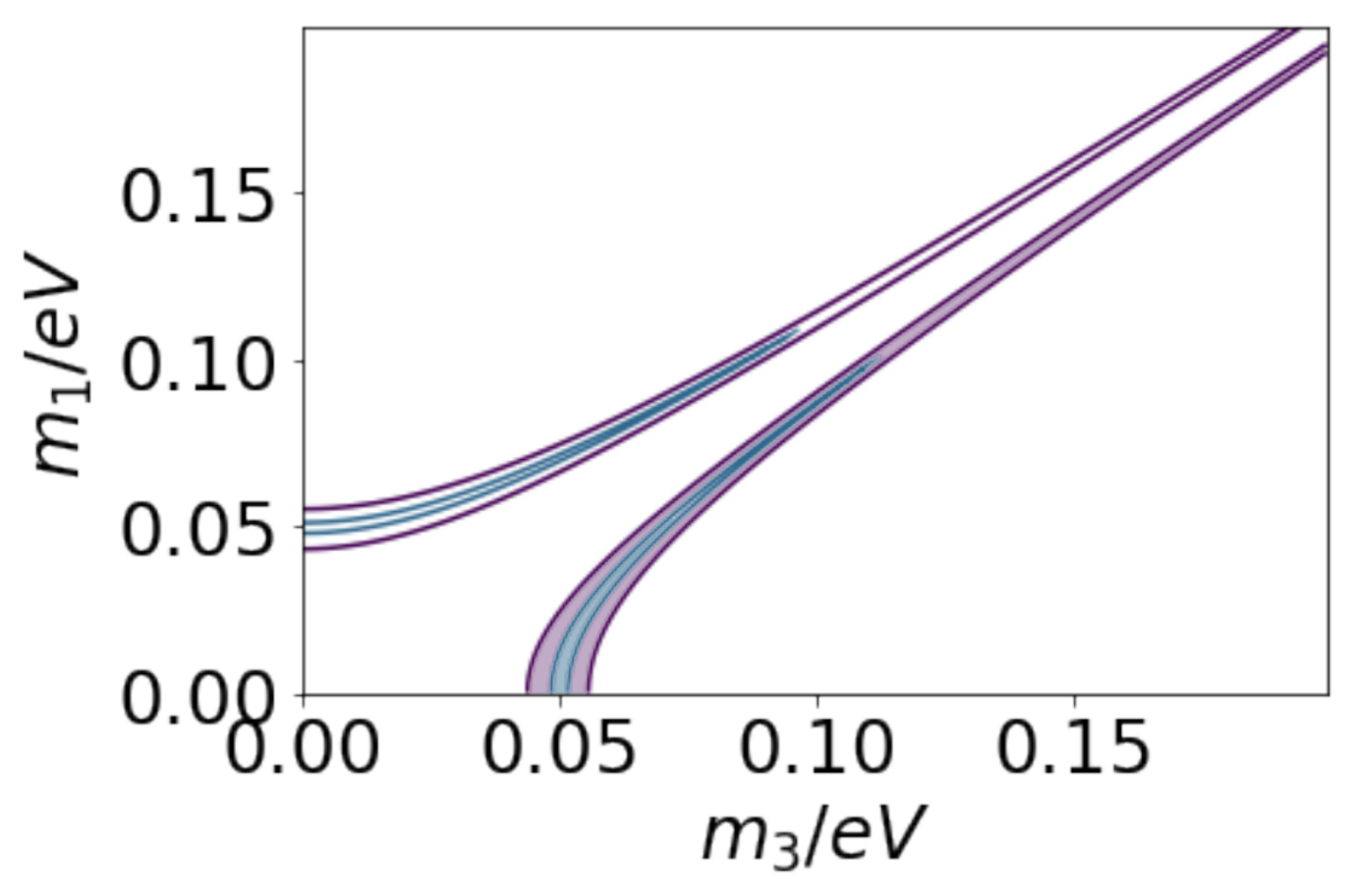}
\caption{Contours of equal posterior probability density for $m_1$ and $m_3$, marginalised over $m_2$, using the reference prior $J$. These masses have the conventional labelling, with $m_1$ being the dominant contributor to the electron neutrino. NH is depicted in solid contours at the bottom right (with $m_1$ identified as $m_{\rm L}$) and IH at top left (with  open contours and $m_1$=$m_{\rm M}$). Contours are set at a very low level to render the posterior visible across the mass range.}
\label{pm1m3}
\end{figure}

For illustration, the integrands marginalised over $m_2$ are shown in figure~\ref{pm1m3} for the two hierarchies.  To compute the Bayes factor, the integrals are taken over the wedge $m_{\rm L} \le m_{\rm M} \le m_{\rm H} \le0.5$ eV.  The upper bound is necessary to make the prior proper, but the results are very insensitive to the choice if sufficiently large, because of the requirement from oscillation data that the masses be almost degenerate if large, and the cosmological bound strongly disfavours $\Sigma=1.5$ eV (i.e. the posterior converges quickly). Furthermore, since the prior volumes are the same for both models, there is essentially no sensitivity of the posterior odds to the chosen prior volume. 

When these marginal likelihoods are evaluated, we find that the posterior odds ratio is
\begin{equation}
\frac{p({\rm NH | \fatt})}{p({\rm IH | \fatt})} = \frac{Z_{\rm NH}}{Z_{\rm IH}} = 5.1
\end{equation}
in mild favour of the normal hierarchy and with a negligible error from the integration.  The result can either be obtained using the $J$ prior and integrating over masses $m_i$, or by using a constant prior and integrating over ($\phi,\psi,\Sigma$), paying attention to the physical domain in the latter case. As expected, the result depends on the cosmological error; for example, increasing the error to $\sigma_\Sigma = 0.07$ eV gives a significantly lower odds ratio of 4.0.   Thus we find, in agreement with \cite{HS2016} and \cite{Vagnozzi}, but in contrast to \cite{SJPV} and \cite{Long}, that the reference prior approach still favours the normal hierarchy, but only with low and inconclusive odds.

This makes mathematically robust what could already intuitively be expected: if multiple reanalyses of the same data yield strongly differing posterior odds, and the difference arises from the choice of priors, then it is clear that all additional information must have come from the prior rather than the data. Here, we have shown this by explicitly using a prior that contributes on average minimal prior information to the information inherent in the data, and we indeed find that using such an uninformative prior leads to uninformative posterior odds. This provides a strong case for collecting better data, in order to determine the neutrino hierarchy. Note, that such an improved data set might still reveal that neutrinos obey the inverse hierarchy: all statements in the paper at hand are entirely probabilistic, and the posterior odds present the odds given current data. The odds may well tip once a measurement of the neutrino mass sum has been established.

\section{Future neutrino constraints}

\begin{figure}
\centering
\includegraphics[width=8cm]{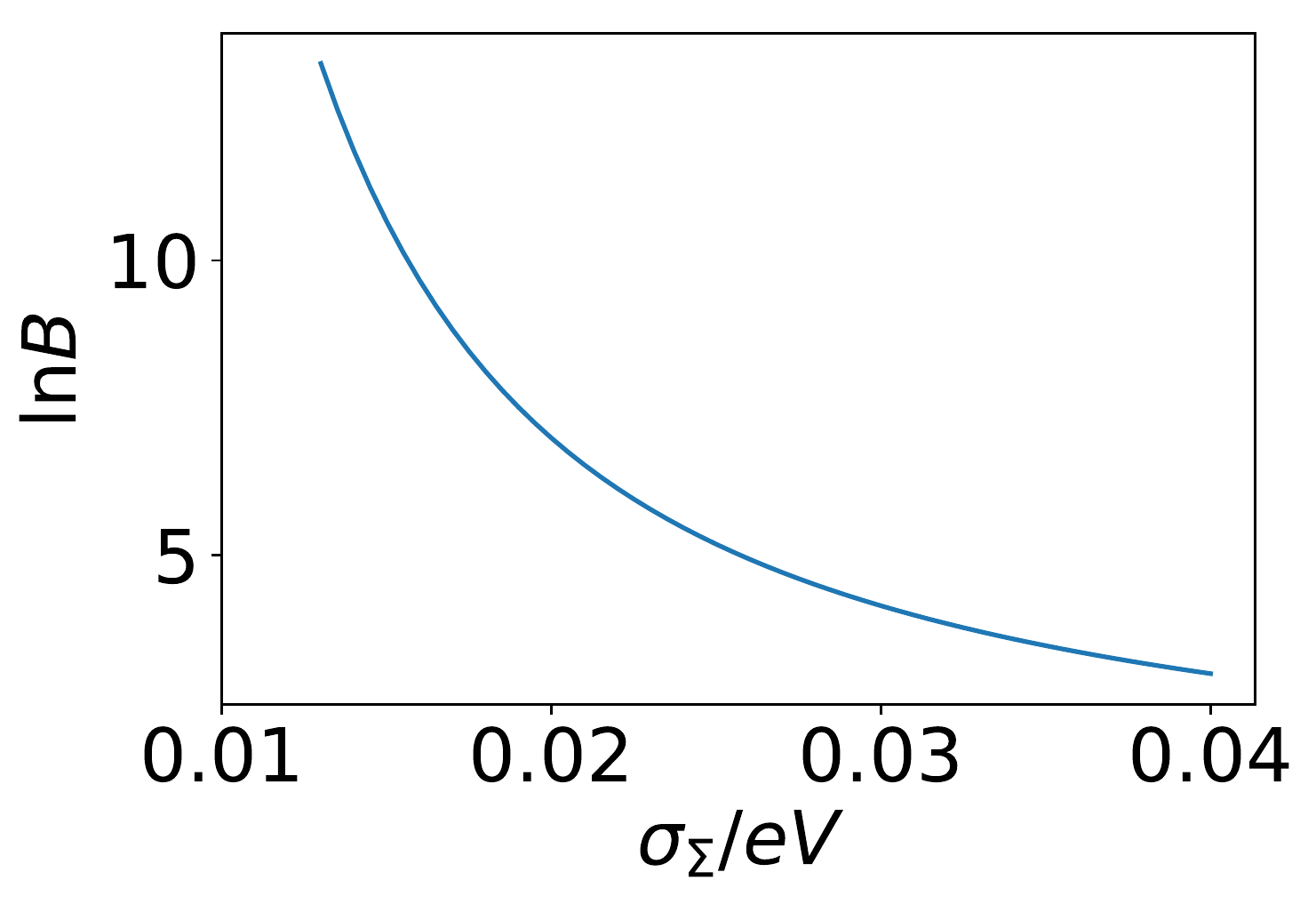}
\caption{The Bayes factor $B=Z_{\rm NH}/Z_{\rm IH}$ for future cosmological experiments where the standard deviation on the sum of neutrino masses reduces from its current value of $0.06$eV. One massless neutrino and the normal hierarchy is assumed.}
\label{futureNH}
\end{figure}

It is straightforward to forecast the Bayes factor for future cosmological experiments, by reducing the variance on the sum of the masses, and centring the cosmological Gaussian on a fiducial sum of masses.  This is shown in figure~\ref{futureNH} for the normal hierarchy as a function of the standard deviation of the assumed Gaussian future error.  We assume that there is one massless neutrino, so for the normal hierarchy we centre the truncated Gaussian on 0.06~eV, and for the inverted hierarchy,  on 0.11~eV.  Assuming the normal hierarchy, we see from the figure that the logarithm of the Bayes factor should exceed 5 for $\sigma_\Sigma<0.025$ eV, similar to that found by \cite{HS2016}.   Future data could yield uncertainties below 0.02 eV with optimistic Cosmic Microwave Background and galaxy clustering measurements \citep{Allison}, and 3D weak lensing from a Euclid-like survey may give errors $\sigma_\Sigma \simeq 0.03$ eV if the masses are not too low \citep{KHVSM}, with the advantage of being insensitive to galaxy bias, and there is the prospect of smaller errors in combination with CMB lensing \cite{KHD}.  These results should be achievable by the mid-2020s, on the same timescale as laboratory experiments are expected to yield evidence of the hierarchy (see \cite{QianVogel} for a review of these prospects).  Improved measurements would be inconclusive if the inverted hierarchy is correct, since the normal hierarchy can increase the minimum mass to find acceptable solutions, and the Bayes factor is always close to unity for very small errors.

\section{Conclusions}

This paper is an objective Bayesian study analysing the neutrino masses and their hierarchy, using summary information from solar, long baseline accelerator, reactor, and cosmological experiments. Since it was found in former studies that the prior still impacts the physical conclusions strongly, we have determined the on average least informative prior for the described neutrino experiments. This is called a reference prior and for the case at hand coincides with the product of one-dimensional Jeffreys priors. In multi-parameter models, the procedure that leads to the reference prior can depend on the ordering of the parameters \citep{Bernardo2005}, but in the case considered here the summary data separate in the $\phi,\psi,\Sigma$ parametrisation, and the reference prior is independent of ordering and is unique, given by eq.~\ref{refprior}.  This prior favours larger masses, which is uninformative with respect to the parameters, but reflects the measurement sensitivity because a given variance in $\phi$ or $\psi$ corresponds to a smaller uncertainty in mass difference if the masses are larger; see Sect.~\ref{ThePrior}. The uninformative prior is shown in figure \ref{PlotMassPrior}. 

Given this prior, the inverted hierarchy is only mildly disfavoured by current data, despite the minimum total mass being close to the cosmological bound.  Using the least informative prior, the posterior odds ratio (assuming equal model priors) is inconclusive, being 5.1:1 in favour of the normal hierarchy.  

As has been remarked upon previously, the conclusions on the neutrino hierarchy are prior-dependent. The results using the non-informative reference prior lie between the low odds ratios of \cite{HS2016,Gerbino,Vagnozzi}, based on uniform mass priors, and the conclusive results of \cite{SJPV,Long} which use the logarithm of the masses as parameters for the prior, and a random mass matrix prior respectively.  The prospects of future cosmological experiments determining the hierarchy depends on which hierarchy it is. With errors on the sum of masses of the order of 0.025 eV or better, they would find strong evidence ($\ln B>5$) for the normal hierarchy if it is true and the smallest mass is negligibly small.  If, on the other hand, the inverted hierarchy is correct, future improvements in cosmological data will not conclusively determine the hierarchy ($\ln B$ does not vary significantly from zero), since the NH can accommodate the larger mass sum by increasing the smallest mass.  

There are a number of refinements that could be made to this first objective Bayesian approach, principally in refining the reference priors to take full account of the small variation in sensitivity of experiments to the mass sum and mass-squared differences.  However, over the range where the likelihood is substantial, the approximation is very good for the terrestrial experiments, and reasonable for the cosmological experiment, so it is difficult to see how the results could change substantially.  Secondly, from a Bayesian perspective a marginalisation over parameters could be done in the precursor LBL/reactor analysis, rather than a profile likelihood that is used to construct the mass-squared difference posteriors \citep{Nufit}. Finally, any dependence of the solar and accelerator-reactor results on each other should be treated by a joint likelihood, but the dependences are very weak, so again, it seems unlikely that residual correlations could turn the inconclusive results into a strong preference for the normal hierarchy, or favour the inverted hierarchy. 

In total, better data are needed before the neutrino masses and their hierarchy can be well determined. The prior here constructed is then still applicable, and stays on average the prior that is most easily overwritten by data.

\acknowledgments

We are grateful to Fergus Simpson, Roberto Trotta, Morgan Wascko, Yoshi Uchida, Raul Jimenez, Licia Verde and Carlos Pe\~ na-Garay for helpful discussions and comments on the manuscript.


\label{lastpage} 
\end{document}